\documentclass[11pt]{article}
\usepackage{amsmath} 
\usepackage{mathtools} 
\usepackage{geometry} 
\newcommand{\half}{{\textstyle \frac{1}{2}}}
\newcommand{\threehalf}{{\textstyle \frac{3}{2}}}
\newcommand{\fivehalf}{{\textstyle \frac{5}{2}}}
\newcommand{\onesixth}{{\textstyle \frac{1}{6}}}

\begin{document}
\title{Eigenmode analysis of the damped Jaynes-Cummings model}

\author {L.G. Suttorp\footnote{email: l.g.suttorp2@contact.uva.nl} \\ 
Institute of Physics, University of  
Amsterdam, \\ Science Park 904, NL-1098 XH Amsterdam, The Netherlands }

\maketitle

\begin{abstract}
The generating functions for density matrix elements of the Jaynes-Cummings
  model with cavity damping are analysed in terms of their eigenmodes, which are
  characterised by a specific temporal behaviour. These eigenmodes are shown to be
  proportional to particular generalised hypergeometric functions. The relative weights of
  these eigenmodes in the generating functions are determined by the initial conditions of
  the model. These weights are found by deriving orthogonality relations involving adjoint
  modes. In an example it is shown how the time-dependent density matrix elements and
  the related factorial moments can be extracted from the eigenmode decompositions
  of the generating functions.
\end{abstract}

  \section{Introduction}\label{intro}
The model introduced by Jaynes and Cummings \cite{JC63} in 1963 continues to draw
attention, as is illustrated by the publication of a collection of papers on the occasion
of its 50th anniversary \cite{GKL13}. The original model, which describes the interaction
of a two-state atom with photons in a cavity mode, has been extended in several ways. In
particular, interesting phenomena show up when damping effects by the escape of photons
from the cavity are included. To incorporate these cavity damping effects a master
equation approach has frequently been employed. It follows by supplementing the equation
governing the time dependence of the density operator with Lindblad terms
\cite{L76}-\cite{GKS76}. Other types of master equations for the damped Jaynes-Cummings
model have been studied recently as well \cite{SMMPM07}-\cite{WN22}.

Various techniques have been employed to solve the master equation for the damped
Jaynes-Cummings model in the Lindblad form. Solutions have been obtained by using
quasi-probability distributions \cite{ER89}-\cite{DRS93} or damping bases \cite{BE93}, or
by starting from the coupled equations for the density operator matrix elements
\cite{vW97}. All of these methods lead to rather complicated expressions for the
time-dependent matrix elements of the density operator. Simpler results have been derived
by making various assumptions about the relative magnitude of the parameters in the model
and the initial form of the density operator \cite{GB93}-\cite{BJ07}.

In \cite{LFQRS01} an attempt has been made to simplify matters by making use of suitable
special functions. It is stated in that paper that the results for the density operator of
the Jaynes-Cummings model with damping can not be fitted to its initial value in a
rigorous way, the reason being that no orthogonality relations are said to be available
for the relevant special functions. In the following, however, we shall obtain the eigenmodes of
the generating functions for the density operator matrix elements of the damped
Jaynes-Cummings model in terms of generalised hypergeometric functions and derive suitable
orthogonality relations for these functions. In this way it will be demonstrated that the
full time dependence of the density operator can be derived, with an exact fitting to the
initial conditions.

\section{Jaynes-Cummings model with cavity damping}
The Lindblad master equation that governs the time evolution of the density operator
$\rho$ for the Jaynes-Cummings model at resonance and with cavity damping reads:
\begin{equation}
  \frac{\partial\rho}{\partial t}=-i [H,\rho]+\kappa (2a\rho a^\dagger -a^\dagger
  a\rho-\rho a^\dagger a) \label{2.1}
\end{equation}
with $a$ and $a^\dagger$ the annihilation and creation operators of the field mode with
frequency $\omega_0>0$, and $\kappa>0$ the damping rate. The Hamiltonian $H$ is:
\begin{equation}
H=\half \omega_0 (|e\rangle\langle e|-|g\rangle\langle g|)+
     \omega_0 a^\dagger  a 
     +f(a|e\rangle\langle g|+ a^\dagger |g\rangle\langle  e|) \label{2.2}
\end{equation}
with $|g\rangle$ and $|e\rangle$ denoting the atomic ground and excited states,
respectively, and $f>0$ the coupling constant. From the master equation one may derive
coupled differential equations for the matrix elements of $\rho$ on the basis of the
states $|g,n\rangle$ and $|e,n\rangle$ of atom and field, with $n$ the number of photons
in the field mode.  For arbitrary $n$ and fixed values of $m-n$ the equations couple the
time evolution of the matrix elements $\langle g,n|\rho|g,m\rangle$,
$\langle g, n+1|\rho|e,m\rangle$, $\langle e,n|\rho| g,m+1\rangle$ and
$\langle e,n|\rho|e,m\rangle$. In the following we will concentrate on the case $m=n$.

Upon introducing the abbreviations
$g_n(\tau)=
\langle g,n|\rho(\tau)|g,n\rangle$, $e_n(\tau)=\langle e,n|\rho(\tau)|e,n\rangle$,
$f_n(\tau)=2\sqrt{n+1}\, {\rm Re} [\langle e,n|\rho(\tau)| g,n+1\rangle]$ and
$h_n(\tau)=2\sqrt{n+1}\, {\rm Im} [\langle e,n|\rho(\tau)| g,n+1\rangle]$ one arrives at a set of
differential equations for $e_n(\tau)$, $f_n(\tau)$, $g_n(\tau)$, and $h_n(\tau)$. It
turns out that these equations simplify by introducing instead of $g_n$ the combinations
$d_n=g_n+e_{n-1}$ for $n>0$, and $d_0=g_0$.  With the scaled variables $\tau=\kappa t$ and
$\alpha=f/\kappa$ we get for $n\geq 0$:
\begin{eqnarray}
 &&   \frac{{\rm d}}{{\rm d}\tau} d_n=2(n+1) d_{n+1}-2nd_n-2 e_n+2e_{n-1}\, , \label{2.3} \\
  &&  \frac{{\rm d}}{{\rm d}\tau} e_n=2(n+1)e_{n+1}-2n e_n-\alpha h_n\, ,\label{2.4} \\
  &&  \frac{{\rm d}}{{\rm d}\tau}f_n=2(n+1) f_{n+1}-(2n+1) f_n\, ,\label{2.5}\\
  &&  \frac{{\rm d}}{{\rm d}\tau}h_n=2(n+1)h_{n+1}-(2n+1)h_n
                         -2\alpha(n+1)d_{n+1}+4\alpha(n+1)e_n\, \label{2.6}.
\end{eqnarray}
In the first equation the last term should be omitted for $n=0$.  The first and the third
equation do not contain the coupling constant $\alpha$. Furthermore, the equation for
$f_n$ is decoupled from those for $d_n$, $e_n$ and $h_n$.

To solve the coupled differential equations (\ref{2.3})-(\ref{2.6}) we introduce the
generating functions $D(z,\tau)=\sum_{n=0}^\infty z^n d_n(\tau)$ and similarly $E(z,\tau)$,
$F(z,\tau)$ and $H(z,\tau)$. The time evolution of these functions is determined by a set
of partial differential equations that follow from (\ref{2.3})--(\ref{2.6}) as
\begin{eqnarray}
  && \frac{\partial D}{\partial\tau}=2(1-z)\frac{\partial D}{\partial z}-2(1-z)E\, , \label{2.7}\\
  && \frac{\partial E}{\partial\tau}=2(1-z)\frac{\partial E}{\partial z}-\alpha H\,
                                     , \label{2.8}\\
  && \frac{\partial F}{\partial\tau}=2(1-z)\frac{\partial F}{\partial z}-F\,
                                     , \label{2.9}\\
  && \frac{\partial H}{\partial\tau}=2(1-z)\frac{\partial H}{\partial z}-H
 -2\alpha \frac{\partial D}{\partial z}+4\alpha \frac{\partial(zE)}{\partial z}\,
  . \label{2.10}
\end{eqnarray}
The function $D(z,\tau)$ is determined up to an additive constant, since only its
derivatives appear in the equations. 

In the following the differential equations (\ref{2.7})-(\ref{2.10}) will be solved in
terms of eigenmodes.  It should be noted that the equation (\ref{2.9}) for $F(z,\tau)$ is
decoupled from those for the other three functions. Although it can easily be solved
directly, it will be analysed in terms of eigenmodes as well, so as to preserve
the analogy in the treatment of the four equations.

\section{Eigenmode solutions}
The eigenmode solution of equation (\ref{2.9}) wil be discussed first. The form of the
equations (\ref{2.7})-(\ref{2.10}) suggests a change of variable from $z$ to $u=1-z$, so
that (\ref{2.9}) becomes an equation for $\bar{F}(u,\tau)=F(1-u,\tau)$.  An equation for
its eigenmodes is obtained by writing $F(u,\tau)=F_\lambda(u){\rm e}^{\lambda\tau}$, where
we suppressed the bar above $F$ again. The resulting eigenmode equation gets the form:
\begin{equation}
  -2u\frac{{\rm d}F_\lambda(u)}{{\rm d}u}-F_\lambda(u)=\lambda F_\lambda(u). \label{3.1}
  \end{equation}
  When the generating function $F(u,\tau)$ is taken to be regular in the interval
  $0 \leq u \leq 1$, one may expand $F_\lambda(u)$ as $\sum_{n=0}^\infty c_n u^n$ near
  $u=0$. Substituting this form in (\ref{3.1}) one finds $(\lambda+2n+1)c_n=0$ for all
  $n\geq 0$, so that a non-trivial solution $F_\lambda(u)=u^k$ is obtained for
  $\lambda=-2k-1$, with non-negative integer $k$.  Upon choosing $k$ as a new label, the
  set of solutions is found as
\begin{equation}
  F_k(u)=u^k  \label{3.2}
  \end{equation}
with non-negative integer $k$. The generating function $F(u,\tau)$ gets the form 
\begin{equation}
  F(u,\tau)=\sum_{k=0}^\infty A_k F_k(u) {\rm e}^{-(2k+1)\tau} \label{3.3}
\end{equation}
in terms of its eigenmodes.  The coefficients $A_k$ can be found from the initial
conditions, as will be shown in the following section.

Next, the eigenmode equations that follow from the coupled partial differential equations for
$D(u,\tau)$, $E(u,\tau)$ and $H(u,\tau)$ will be considered. Upon changing variables from
$z$ to $u$ in (\ref{2.7}), (\ref{2.8}) and (\ref{2.10}), and assuming an exponential time
dependence as before, one gets:
\begin{eqnarray}
&& -2u\frac{{\rm d}D_\lambda(u)}{{\rm d}u}-2uE_\lambda(u)=\lambda D_\lambda(u)\, , \label{3.4}\\
 && -2u\frac{{\rm d}E_\lambda(u)}{{\rm d}u}-\alpha H_\lambda(u)=\lambda E_\lambda(u)\, ,\label{3.5}\\
 && -2u\frac{{\rm d}H_\lambda(u)}{{\rm d}u}-H_\lambda(u)+2\alpha\frac{{\rm
                                             d}D_\lambda(u)}{{\rm d}u}
-4\alpha(1-u)\frac{{\rm d}E_\lambda}{{\rm d}u}
  +4\alpha E_\lambda(u)=\lambda H_\lambda(u)\, . \label{3.6}
\end{eqnarray}
After eliminating $D_\lambda(u)$ and $H_\lambda(u)$, one arrives at a third-order differential
equation for $E_\lambda(u)$:
\begin{eqnarray}
  &&8u^3\frac{{\rm d}^3E_\lambda(u)}{{\rm
     d}u^3}+4u\left[u(3\lambda+2a+9)-2a\right]\frac{{\rm d}^2E_\lambda(u)}{{\rm d}u^2}
+2\left[u(3\lambda^2+2a\lambda+12\lambda+10a+12)\right.\nonumber\\
     &&\hspace{5mm}\left. -2a\lambda-4a\right]\frac{{\rm
     d}E_\lambda(u)}{{\rm d}u}+(\lambda^3+3 \lambda^2+4a\lambda+2\lambda+4a)E_\lambda(u)=0\,
  \label{3.7}
  \end{eqnarray}
with $a=\alpha^2$. Insertion of a series of the form $E_\lambda(u)=\sum_{n=0}^\infty c_n
u^n$ leads to a recursion relation for the coefficients:
\begin{equation}
  (n+1)(n+\half\lambda+1)a\, c_{n+1}
  =(n+\half\lambda+\half) (n+\half\lambda+\half a+\half
     w_\lambda+\half) 
   (n+\half\lambda+\half a-\half w_\lambda+\half)\, 
     c_n  \label{3.8}
\end{equation}
with the abbreviation $w_\lambda=\sqrt{(a-1)^2+2a\lambda}$.  For general values of
$\lambda$ the series representing $E_\lambda(u)$ diverges near $u= 0$. A convergent result
is found only if the series terminates after a finite number of terms. This may happen in
several different ways: either one has $\half\lambda+\half=-k$ or
$\half\lambda+\half a \mp \half w_\lambda+\half=-k$, with non-negative integer $k$.

In the first case, for $\lambda=-2k-1$, one has $w_\lambda=\sqrt{a^2-4a(k+1)+1}\equiv
w_k$. The solution for $E_\lambda(u)$ is proportional to a terminating generalised
hypergeometric function $\mbox{}_3F_1$:
\begin{equation}
  \mbox{}_3F_1(-k,-k+\half a+\half w_k, -k+\half a-\half
  w_k;-k+\half;\frac{u}{a}) \, . \label{3.9}
  \end{equation}
  By inverting the order of the terms in the finite series one may write the functions
  $E_{0,k}(u)$ of this first set of eigenmodes in terms of terminating generalised
  hypergeometric functions $\mbox{}_2F_2$:
\begin{equation}
 E_{0,k}(u)=\left(\frac{u}{a}\right)^k\, \mbox{}_2F_2(-k,\half;
  -\half a+1+\half w_k,-\half a+1-\half w_k;-\frac{a}{u}) \, . \label{3.10}
\end{equation}
The functions $D_{0,k}(u)$ and $H_{0,k}(u)$ of these eigenmodes are obtained from
(\ref{3.4})--(\ref{3.6}) as
\begin{eqnarray}
 && D_{0,k}(u)=-2a\left(\frac{u}{a}\right)^{k+1}\, \mbox{}_2F_2(-k,-\half ; 
            -\half a+1+\half w_k,-\half a+1-\half w_k;-\frac{a}{u})\, , \label{3.11}\\
 && H_{0,k}(u)=\frac{1}{\sqrt{a}}\left(\frac{u}{a}\right)^k \,
                \mbox{}_2F_2(-k,\threehalf;
            -\half a+1+\half w_k,-\half a+1-\half w_k;-\frac{a}{u})\, . \label{3.12}
               \end{eqnarray}

In the other two cases, with $\half\lambda+\half a \mp \half w_\lambda+\half=-k,$ the
eigenvalues get the form $\lambda=-2k-1\pm \bar{w}_k$ with $\bar{w}_k\equiv
\sqrt{1-4a(k+1)}$. Again, the solution for $E_\lambda(u)$ is proportional to a terminating generalised
hypergeometric function:
\begin{equation}
 \mbox{}_3F_1(-k,-k\pm\half\bar{w}_k, -k+a\pm\half
    \bar{w}_k; -k+\half\pm\half\bar{w}_k;\frac{u}{a})\, . \label{3.13}
 \end{equation}
As before, the order of the terms can be inverted, with the result
\begin{equation}
  E_{\pm,k}(u)= \left(\frac{u}{a}\right)^k \,
                  \mbox{}_2F_2(-k,\half\mp\half \bar{w}_k;
  1\mp\half\bar{w}_k,1-a\mp\bar{w}_k;-\frac{a}{u}) \, .\label{3.14}
\end{equation}
The associated functions $D_{\pm,k}(u)$ and $H_{\pm,k}(u)$ are
\begin{eqnarray}
  && D_{\pm,k}(u)=-\frac{2a}{1\pm\bar{w}_k}\left(\frac{u}{a}\right)^{k+1}\,
                  \mbox{}_2F_2(-k,-\half\mp\half\bar{w}_k;
              1\mp\half\bar{w}_k,1-a\mp\bar{w}_k;-\frac{a}{u})\, , \label{3.15}\\
  && H_{\pm,k}(u)=\frac{1}{\sqrt{a}}(1\mp\bar{w}_k)\left(\frac{u}{a}\right)^k\,
                  \mbox{}_2F_2(-k,\threehalf\mp\half\bar{w}_k;
              1\mp\half\bar{w}_k,1-a\mp\bar{w}_k;-\frac{a}{u}) \, .
                 \label{3.16}
                 \end{eqnarray}
                 
The generating function $E(u,\tau)$ can be expanded in terms of the
eigenmodes (\ref{3.10}) and (\ref{3.14}):
\begin{equation}
E(u,\tau)=\sum_{s=0,\pm}\sum_{k=0}^\infty A_{s,k} E_{s,k}(u){\rm
e}^{(-2k-1+s\bar{w}_k)\tau}\label{3.17}
\end{equation}
with coefficients $A_{s,k}$ that follow from the initial conditions. The generating
functions $D(u,\tau)$ and $H(u,\tau)$ get analogous forms:
\begin{eqnarray}
&&D(u,\tau)=\sum_{s=0,\pm}\sum_{k=0}^\infty A_{s,k} D_{s,k}(u)
   {\rm e}^{(-2k-1+s\bar{w}_k)\tau} 
+1\, , \label{3.18}\\
&&H(u,\tau)=\sum_{s=0,\pm}\sum_{k=0}^\infty A_{s,k} H_{s,k}(u)
             {\rm e}^{(-2k-1+s\bar{w}_k)\tau}\, . \label{3.19}
\end{eqnarray}
The generating function $G(u,\tau)$ follows from $E(u,\tau)$ and $D(u,\tau)$ as
\begin{equation}
  G(u,\tau)=D(u,\tau)-(1-u)E(u,\tau) \label{3.20}
\end{equation}
since $g_n$ equals $d_n-e_{n-1}$ for $n>0$ and $d_0=g_0$.  For $u=0$ the relation
(\ref{3.20}) implies $G(0,\tau)+E(0,\tau)=D(0,\tau)$. Because the normalisation of the
density operator implies $\sum_{n=0}^\infty [g_n(\tau)+e_n(\tau)]=1$ for all $\tau$, the
function $D(u,\tau)$ should equal 1 for $u=0$. For that reason a constant term has been
added to the right-hand side of  (\ref{3.18}).

The function $E(u,\tau)$ depending on $u$ is the generating function
of the factorial moments $\bar{e}_m$ associated to $e_n$. In fact, from the relation
$u=1-z$ it follows that the definition $E(z,\tau)=\sum_{n=0}^\infty z^ne_n(\tau)$ leads to
the expansion
\begin{equation}
  E(u,\tau)=\sum_{m=0}^\infty \frac{(-1)^m}{m!} u^m\, \bar{e}_m(\tau) \label{3.21}
\end{equation}
with the factorial moments defined as
\begin{equation}
  \bar{e}_m(\tau)=\sum_{n=m}^\infty \frac{n!}{(n-m)!}\, e_n(\tau)\, . \label{3.22}
\end{equation}
The lowest-order factorial moment is $\bar{e}_0(\tau)=\sum_{n=0}^\infty e_n(\tau)$. It is
obtained from $E(u,\tau)$ by putting $u$ equal to 0. The functions $E_{s,k}(u)$ in
(\ref{3.17}) are polynomials in $u$, which are finite at $u=0$. The expansion of
$G(u,\tau)$ in factorial moments $\bar{g}_m(\tau)$ is similar to (\ref{3.21}) with
(\ref{3.22}). The normalisation of the density operator can be written in terms of the
factorial moments as $\bar{g}_0(\tau)+\bar{e}_0(\tau)=1$ for all $\tau$.

The expressions (\ref{3.10}-(\ref{3.12}) and (\ref{3.14})-(\ref{3.16}) for the eigenmodes
can easily be rewritten as polynomials in $z=1-u$. One  finds for instance from
(\ref{3.10}) with the help of the binomial theorem:
\begin{equation}
  E_{0,k}(z)=\frac{k!}{a^k}\sum_{n=0}^k \frac{(-1)^n}{n!(k-n)!} \, \mbox{}_2F_2(-k+n,\half;
  -\half a+1+\half w_k,-\half a+1-\half w_k;-a) z^n\, .  \label{3.23}
\end{equation}
As the dependence on $z$ is now made explicit the contribution of the eigenmode with label
$0,k$ to the density matrix element $e_n$ follows directly.
  
The results (\ref{3.17})-(\ref{3.19}) with (\ref{3.10})-(\ref{3.12}) and
(\ref{3.14})-(\ref{3.16}) give the complete eigenmode expansions of the generating
functions in terms of generalised hypergeometric functions. They contain coefficients 
$A_{s,k}$, which still have to be determined.  
                 
\section{Adjoint modes}
The relative weights $A_{s,k}$ in the eigenmode expansions may be obtained from the
initial conditions at $\tau=0$. Their form can be found by considering the adjoint
differential equations and their eigenmodes. The solutions (\ref{3.10})--(\ref{3.12}) and
(\ref{3.14})--(\ref{3.16}) suggest that it is convenient to choose the new independent
variable $x=a/u$ instead of $u$. For analogy, we shall use the variable $x$ to determine
the adjoint eigenmodes associated to (\ref{3.2}) as well, although its simple form does
not point in that direction.

In terms of $x$ the eigenmode equation (\ref{3.1}) reads:
\begin{equation}
  2x\frac{{\rm d}F_\lambda(x)}{{\rm d}x} -F_\lambda(x)=\lambda F_\lambda(x)\, . \label{4.1}
  \end{equation}
  Its adjoint equation is
  \begin{equation}
  - 2x\frac{{\rm d}\hat{F}_\lambda(x)}{{\rm d}x} -3\hat{F}_\lambda(x)=\lambda \hat{F}_\lambda(x)\,  \label{4.2}
  \end{equation}
with the solution $\hat{F}_\lambda(x)=x^{-(\lambda+3)/2}$ up to a constant factor. Upon choosing the same
eigenvalue spectrum as in (\ref{3.2}) by writing $\lambda=-2m-1$ with non-negative integer
$m$, we write the adjoint eigenmodes as $\hat{F}_m(x)=c_mx^{m-1}$.

The eigenmodes $F_k(x)$ and their adjoints $\hat{F}_m(x)$ satisfy an orthogonality
relation involving a contour integral in the complex $x$-plane around $x=0$:
\begin{equation}
\frac{1}{2\pi {\rm i}}\oint {\rm d}x \, \hat{F}_m(x) F_k(x)=\delta_{m,k} \label{4.3}
\end{equation}
if $c_m$ is chosen to be equal to $a^{-m}$.  This identity may be proven directly by
substituting the explicit expressions for $\hat{F}_m(x)$ and $F_k(x)$. A formal proof for
$m\neq k$ starts by evaluating the integral $\oint {\rm d}x \, \hat{F}_m(x)$
$ [2x\, {\rm d} F_k(x)/{\rm d} x-F_k(x)]$ in two ways, either by employing (\ref{4.1}) or
by using (\ref{4.2}) after an integration by parts. Equating the two results one arrives
at (\ref{4.3}) for $m\neq k$. The orthogonality relation (\ref{4.3}) can be employed to
determine the coefficients $A_k$ in (\ref{3.3}) as:
\begin{equation}
  A_k=\frac{1}{2\pi{\rm i}}\oint {\rm d}x \, \hat{F}_k(x)\, 
               F(x,0) \, .\label{4.4}
 \end{equation}

 After this rather simple case we now turn to the coupled set of equations (\ref{2.7}),
 (\ref{2.8}) and (\ref{2.10}). They have led to the eigenmode equations
 (\ref{3.4})--(\ref{3.6}).  Rewriting these equations in terms of $x$ we get
\begin{eqnarray}
&&2x\frac{{\rm d}D_\lambda(x)}{{\rm d}x} -2\frac{a}{x} E_\lambda(x)=\lambda D_\lambda(x)\, , \label{4.5}\\
&&2x\frac{{\rm d}E_\lambda(x)}{{\rm d}x}-\sqrt{a} H_\lambda(x) = \lambda E_\lambda(x)\,
, \label{4.6}\\
&&2x\frac{{\rm d}H_\lambda(x)}{{\rm d}x}-H_\lambda(x)-\frac{2}{\sqrt{a}}x^2\frac{{\rm
                                         d}D_\lambda(x)}{{\rm d}x}
+\frac{4}{\sqrt{a}}x(x-a)\frac{{\rm d}E_\lambda(x)}{{\rm d}x}+4\sqrt{a}
     E_\lambda(x)=\lambda H_\lambda(x)\, . \label{4.7}
\end{eqnarray}

The adjoint differential equations are
\begin{eqnarray}
&& -2x\frac{{\rm d}\hat{D}_\lambda(x)}{{\rm d}x}-2\hat{D}_\lambda(x)+\frac{2}{\sqrt{a}}x^2\frac{{\rm d}\hat{H}_\lambda(x)}{{\rm d}x}+\frac{4}{\sqrt{a}}x\hat{H}_\lambda(x)=\lambda \hat{D}_\lambda(x)\, , \label{4.8}\\
&&-2x\frac{{\rm d}\hat{E}_\lambda(x)}{{\rm
   d}x}-2\hat{E}_\lambda(x)-\frac{2a}{x}\hat{D}_\lambda(x)
-\frac{4}{\sqrt{a}}x(x-a)\frac{{\rm d}\hat{H}_\lambda(x)}{{\rm d}x}
-\frac{8}{\sqrt{a}}(x-a)\hat{H}_\lambda(x)=\lambda\hat{E}_\lambda(x)\,, \label{4.9}\\
  &&-2x\frac{{\rm d}\hat{H}_\lambda(x)}{{\rm
     d}x}-3\hat{H}_\lambda(x)-\sqrt{a}\hat{E}_\lambda(x)
=\lambda \hat{H}_\lambda(x)\, . \label{4.10}
  \end{eqnarray}
  Elimination of $\hat{D}_\lambda$ and $\hat{H}_\lambda$ yields the third-order
  differential equation
  \begin{eqnarray}
    &&8x^3 \frac{{\rm d}^3\hat{E}_\lambda(x)}{{\rm d}x^3}+
       4x^2(-2x+3\lambda+2a+13)\frac{{\rm d}^2\hat{E}_\lambda(x)}{{\rm d}x^2}
    +2x\left[-2(\lambda+12)x+3\lambda^2+2a\lambda+20\lambda\right. \nonumber\\
      &&\left.+14a+32\right]\frac{{\rm d}\hat{E}_\lambda(x)}{{\rm d}x}
  +\left[-8(\lambda+6)x+\lambda^3
     +7\lambda^2 +8a\lambda+14\lambda+8a+8\right]
       \hat{E}_\lambda(x)=0\, . 
    \label{4.11}
  \end{eqnarray}
  For arbitrary values of $\lambda$ three independent solutions are
  \begin{eqnarray}
 &&\hat{E}_{0,\lambda}(x)=x^{-(\lambda+1)/2}\, \mbox{}_2F_2(-\half\lambda+\threehalf,\fivehalf;
\half a+2+\half w_\lambda,\half a+2-\half w_\lambda;x)\, , \label{4.12}\\
&&\hat{E}_{\pm,\lambda}(x)=x^{-(\lambda+a+3\mp w_\lambda)/2}\, \mbox{}_2F_2(-\half
                            a+\threehalf\pm \half w_\lambda,
 -\half a-\half\lambda+\half\pm\half w_\lambda;-\half a\pm\half
   w_\lambda,1\pm w_\lambda;x) \hspace{10mm}\label{4.13}
  \end{eqnarray}
  with $w_\lambda=\sqrt{(a-1)^2+2a\lambda}$ as before.

  From the solutions for general $\lambda$ a suitable set of adjoint eigenmodes will be
  obtained by imposing the condition that the spectrum is the same as that found for the eigenmodes
  in the previous section. Hence, one should take either $\lambda=-2m-1$ or
  $\lambda=-2m-1\pm\bar{w}_m$, with non-negative integer $m$. Upon choosing solutions that
  are either analytic or having a simple pole at $x=0$ we get from (\ref{4.12}) with
  $\lambda=-2m-1$:
  \begin{equation}
  \hat{E}_{0,m}(x)=c_{0,m} x^m \mbox{}_2F_2(m+2,\fivehalf ;
\half a+2+\half w_m,\half a+2-\half w_m;x) \label{4.14}
 \end{equation}
 with $w_m=\sqrt{a^2-4a(m+1)+1}$ and with an as yet arbitrary constant $c_{0,m}$.  From
 (\ref{4.8})-(\ref{4.10}) the associated functions $\hat{D}_{0,m}$ and $\hat{H}_{0,m}$ are
 obtained as
 \begin{eqnarray} 
&&\hat{D}_{0,m}(x)= c'_{0,m}\, x^m\, \mbox{}_2F_2(m+2,\half;
\half a+1+\half w_m,\half a+1-\half w_m;x)\, , \label{4.15}\\
  && \hat{H}_{0,m}(x)= c''_{0,m}\, x^{m-1}\, \mbox{}_2F_2(m+1,\threehalf;
 \half a+1+\half w_m,\half a +1-\half w_m;x) \label{4.16}
 \end{eqnarray}
 with the coefficients $c'_{0,m}=-\onesixth(4am+8a+3)\, c_{0,m}$ and $c''_{0,m}=\half [\sqrt{a} /(m+1)] c'_{0,m}$.

 Likewise, we find from (\ref{4.13}) for $\lambda=-2m-1\pm\bar{w}_m$, with $\bar{w}_m=\sqrt{1-4a(m+1)}$:
 \begin{equation}
   \hat{E}_{\pm,m}(x)= c_{\pm,m}\, x^{m-1}\, \mbox{}_2F_2(m+1,\threehalf\pm\half
                         \bar{w}_m;
   \pm\half\bar{w}_m,1+a\pm\bar{w}_m;x)\, . \label{4.17}
 \end{equation}
 The associated functions $\hat{D}_{\pm,m}$ and $\hat{H}_{\pm,m}$ are:
\begin{eqnarray}
  &&\hat{D}_{\pm,m}(x)= c'_{\pm,m}\, x^m\, \mbox{}_2F_2(m+2,\half\pm\half\bar{w}_m;
1\pm\half\bar{w}_m,1+a\pm\bar{w}_m;x)\, , \label{4.18}\\
&&\hat{H}_{\pm,m}(x)= c''_{\pm,m}\, x^{m-1}\,
                        \mbox{}_2F_2(m+1,\threehalf\pm\half\bar{w}_m;
  1\pm\half\bar{w}_m,1+a\pm\bar{w}_m;x) \label{4.19}
\end{eqnarray}
with $c'_{\pm,m}=\mp[(1\mp\bar{w}_m) /(2a\bar{w}_m)] c_{\pm,m}$ and $c''_{\pm,m}=
[2a^{3/2}/(1\mp\bar{w}_m)] c'_{\pm,m}  $.

The eigenmodes (\ref{3.10})--(\ref{3.12}), (\ref{3.14})--(\ref{3.16}) and their adjoints
(\ref{4.14})--(\ref{4.19}) satisfy orthogonality relations of the form:
\begin{equation}
  \frac{1}{2\pi {\rm i}}\oint {\rm d}x \left[ \hat{E}_{r,m}(x)\,
     E_{s,k}(x)+\hat{D}_{r,m}(x)\, D_{s,k}(x)
 + \hat{H}_{r,m}(x)\, H_{s,k}(x)\right]=\delta_{r,s}\, \delta_{k,m} \label{4.20}
\end{equation}
for $r=0,\pm$ and $s=0,\pm$ and for all non-negative integers $k,m$. The contour integral
in the complex $x$-plane encircles the origin $x=0$. The normalisation constants of the
adjoint modes have to be chosen as $c'_{0,m}=2(m+1)/[1-4a(m+1)]$ and
$c'_{\pm,m}=-(m+1)/[1-4a(m+1)]$. The proof of the orthogonality relations for $r\neq s$
and/or $k\neq m$ follows by multiplying the left-hand side of (\ref{4.20}) by the factor
$-2k-1+s\bar{w}_k$, using (\ref{4.5})--(\ref{4.7}), integrating by parts and employing
(\ref{4.8})--(\ref{4.10}), with a result that is again proportional to the left-hand side
of (\ref{4.20}), with a different factor $-2m-1+r\bar{w}_m$, so that the integral must
vanish. For the diagonal case $r=s$ and $k=m$ the relation may be verified by inserting
the explicit forms of the eigenmodes and their adjoints. The result of the integration is
found from the leading terms of the generalised hypergeometric functions.

Once the orthogonality relations have been established we may use them to find the
coefficients $A_{s,k}$ in the expansions (\ref{3.17})--(\ref{3.19}). In fact, upon
changing variables from $u$ to $x=a/u$, putting $\tau=0$, multiplying each of these
expansions by the corresponding expression of an adjoint mode (with fixed parameters $r$
and $m$), summing the results and integrating over $x$ one gets from (\ref{4.20}):
\begin{equation}
 A_{r,m}=\frac{1}{2\pi{\rm i}}\, \oint {\rm d}x\left[\hat{E}_{r,m}(x)\, E(x,0)
  +\hat{D}_{r,m}(x)\, D(x,0) +\hat{H}_{r,m}(x)\, H(x,0)\right]\,  . \label{4.21}
\end{equation} 
Since the functions $\hat{D}_{r,m}(x)$ are analytic in $x=0$ the last term in (\ref{3.18})
does not contribute to $A_{r,m}$.

\section{Example}
As an example of the use of eigenmodes in analysing the behaviour of the damped
Jaynes-Cummings model a special case will be considered. It follows by assuming that at
$\tau=0$ the atom is in its ground state, with $n_0$ photons present. Hence, the initial
value of $d_n$ is given by $d_n(0)=\delta_{n,n_0}$. Furthermore, $e_n(0)$ and $h_n(0)$
vanish for all $n$. The generating function $D(x,0)$ gets the form
\begin{equation}
  D(x,0)=\sum_{p=0}^{n_0} \frac{(-1)^pn_0!}{p!(n_0-p)!}\left(\frac{a}{x}\right)^p \label{5.1}
\end{equation}
while $E(x,0)$ en $H(x,0)$ both vanish. The coefficient $A_{s,k}$ as given by (\ref{4.21})
becomes:
\begin{equation}
  A_{s,k}=\sum_{p=0}^{n_0} \frac{(-1)^pn_0!}{p!(n_0-p)!}\frac{1}{2\pi{\rm i}}\, \oint {\rm
    d}x \left(\frac{a}{x}\right)^p \hat{D}_{s,k}(x)\, . \label{5.2}
\end{equation}
For $s=0$ one gets after substituting (\ref{4.15}) and performing the contour integral
around the origin:
\begin{equation}
  A_{0,k}= c'_{0,k}\frac{(-a)^{k+1} n_0!}{(k+1)!(n_0-k-1)!}\, 
  \mbox{}_2F_2(-n_0+k+1,\half;\half a+1+\half w_k,\half a+1-\half w_k;a) \label{5.3}
\end{equation}
for $0\leq k\leq n_0-1$. Likewise, one obtains
for $s=\pm$ and $0\leq k\leq n_0-1$:
\begin{equation}
  A_{\pm,k}= c'_{\pm,k}\frac{(-a)^{k+1} n_0!}{(k+1)!(n_0-k-1)!}\, 
  \mbox{}_2F_2(-n_0+k+1,
\half  \pm\half\bar{w}_k;1 \pm\half \bar{w}_k,1+a\pm \bar{w}_k;a)\, . \label{5.4}
\end{equation}
The normalisation constants $c'_{0,k}$ and $c'_{\pm,k}$ have been defined below
(\ref{4.20}).

The generating function $E(u,\tau)$ of the factorial moments $\bar{e}_m(\tau)$ follows by
insertion of (\ref{3.10}), (\ref{3.14}) and (\ref{5.3})-(\ref{5.4}) in (\ref{3.17}). The
resulting expression is a sum over $k$ (with $0\leq k\leq n_0-1$) of products of two
terminating generalised hypergeometric functions (one with the argument $a$ and the other
with the argument $-a/u$ and a pre-factor $u^k$), and a time-dependent exponential factor.

\begin{figure}[h]
  \begin{center}
    \includegraphics[height=6cm]{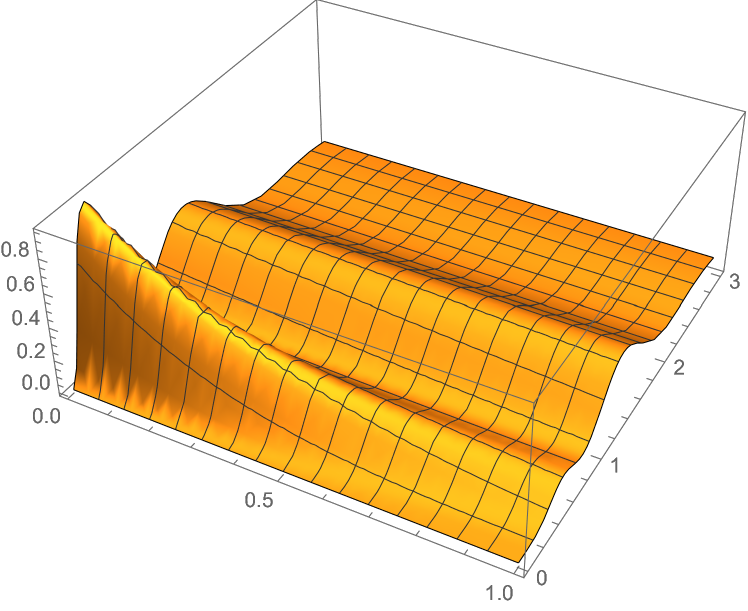}
    \caption{The generating function $E(u,\tau)$ as a function of $u$ (for $0\leq u \leq
    1$) and $\tau$ (for $0\leq \tau \leq 3$), for $n_0=6$ and $a=5$.}
    \label{figure1}
  \end{center}
  \end{figure}

As an illustration, the characteristic wavelike behaviour of the generating function
$E(u,\tau)$ is shown in Figure \ref{figure1} for $n_0=6$ and $a=5$. 

\begin{figure}[h]
  \begin{center}
    \includegraphics[height=5cm]{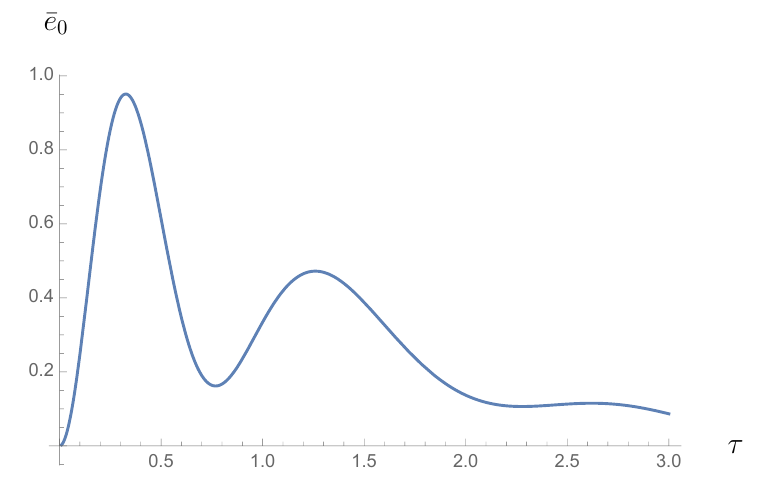}
    \caption{The factorial moment $\bar{e}_0(\tau)$ as a function of $\tau$ (for $0\leq \tau \leq 3$), for $n_0=6$ and $a=5$.}
    \label{figure2}
  \end{center}
\end{figure}

\begin{figure}[b]
  \begin{center}
    \includegraphics[height=5cm]{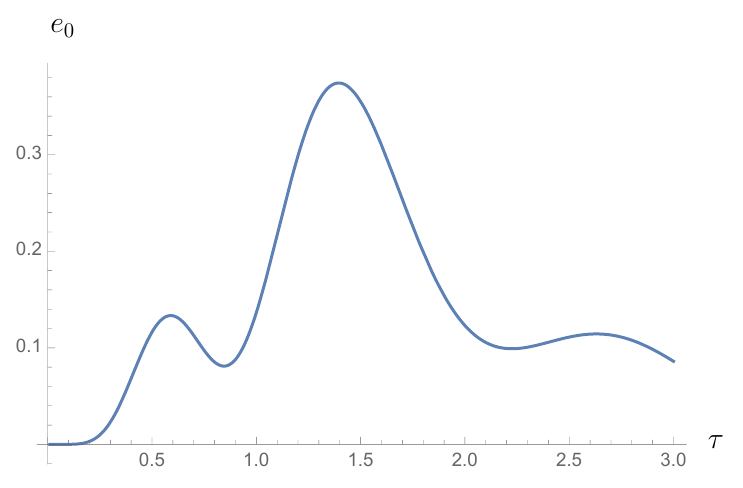}
    \caption{The density matrix element $e_0(\tau)$ as a function of $\tau$ (for $0\leq \tau \leq 3$), for $n_0=6$ and $a=5$.} 
    \label{figure3}
  \end{center}
\end{figure}

  The lowest-order factorial moment $\bar{e}_0(\tau)$ is obtained from $E(u,\tau)$ by
  setting $u$ equal to 0. Its value, as given in Figure \ref{figure2}, determines the
  probability of the atom being in its excited state for any number of photons in the
  cavity. Starting from 0 at $\tau=0$ it returns to that value in the course of time. 
On the other hand, the lowest-order density matrix element $e_0(\tau)$ follows from
$E(u,\tau)$ by taking $u=1$, or $z=0$. As shown in Figure \ref{figure3}, it gives the
probability of finding the atom at time $\tau$ in its excited state and no photons
present. Clearly $e_0(\tau)$ is less than (or equal to) $\bar{e}_0(\tau)$ for all $\tau$.

\begin{figure}[h]
  \begin{center}
    \includegraphics[height=6cm]{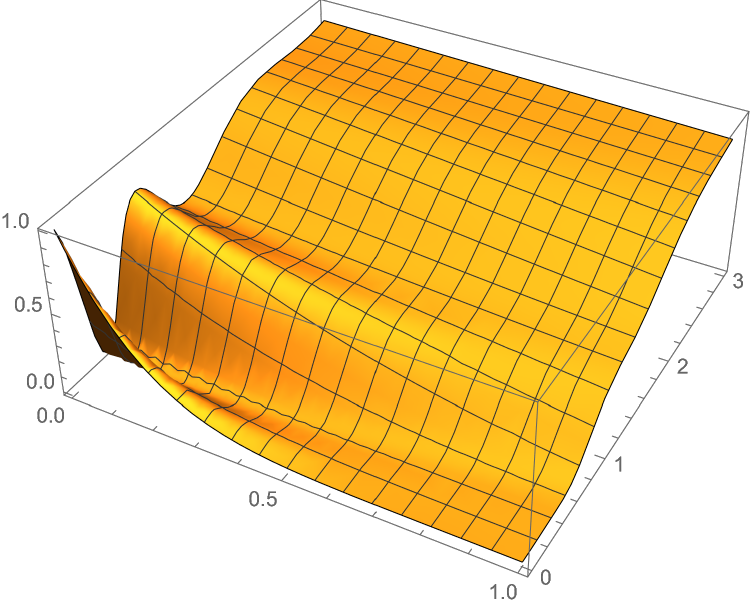}
    \caption{The generating function $G(u,\tau)$ as a function of $u$ (for $0\leq u \leq
    1$) and $\tau$ (for $0\leq \tau \leq 3$), for $n_0=6$ and $a=5$.}
    \label{figure4}
  \end{center}
\end{figure}

Next, we turn to the generating function $G(u,\tau)$ of the density matrix elements
$g_n(\tau)$. It can be found by considering a suitable combination of $E(u,\tau)$ and
$D(u,\tau)$, as given by (\ref{3.20}). The function $D(u,\tau)$ may be obtained by
substitution of (\ref{3.11}), (\ref{3.15}) and (\ref{5.3})-(\ref{5.4}) in (\ref{3.18}).
Combining $E(u,\tau)$ and $D(u,\tau)$, as in (\ref{3.20}), we get an expression for
$G(u,\tau)$. Its behaviour is shown in Figure \ref{figure4} for the same values of $n_0$
and $a$ as above.

\begin{figure}[h]
  \begin{center}
    \includegraphics[height=5cm]{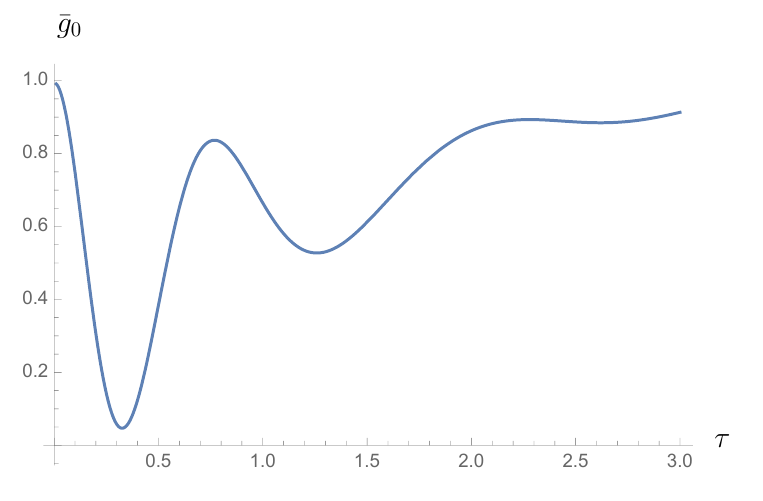}
    \caption{The factorial moment $\bar{g}_0(\tau)$ as a function of $\tau$ (for $0\leq \tau \leq 3$), for $n_0=6$ and $a=5$.}
    \label{figure5}
  \end{center}
\end{figure}

\begin{figure}[b]
  \begin{center}
    \includegraphics[height=5cm]{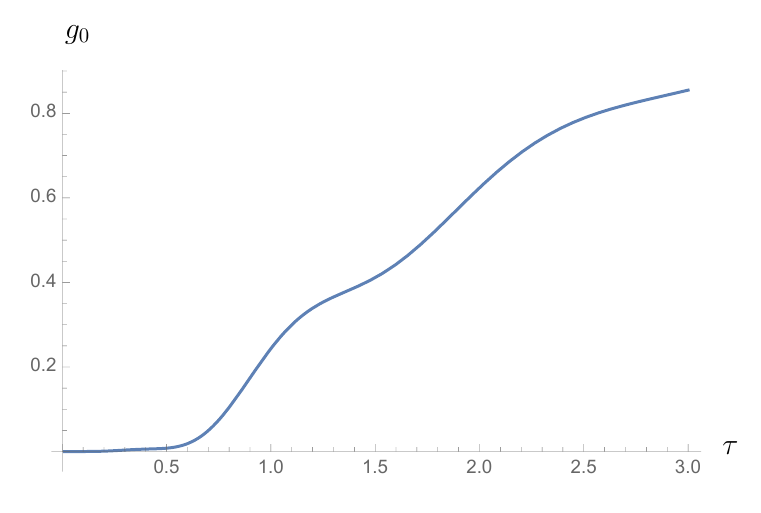}
    \caption{The density matrix element $g_0(\tau)$ as a function of $\tau$ (for $0\leq \tau \leq 3$), for $n_0=6$ and $a=5$.} 
    \label{figure6}
  \end{center}
\end{figure}

Again, one may consider the lowest-order factorial moment
$\bar{g}_0(\tau)$  (see Figure \ref{figure5}). It gives the probability of the
atom being in its ground state regardless of the number of photons present.
Comparison of the Figures \ref{figure2} and \ref{figure5} shows that the two lowest-order
factorial moments $\bar{e}_0(\tau)$ and $\bar{g}_0(\tau)$ add up to 1 for all $\tau$, as expected.
Finally, the time behaviour of the lowest-order density matrix element $g_0(\tau)$ is
shown in Figure \ref{figure6}. It is rising from its initial value 0 to (nearly) its final
value 1 in the time span considered here.  

The expressions for the generating functions that follow by inserting the coefficients
(\ref{5.3})-(\ref{5.4}) in (\ref{3.17})-(\ref{3.19}) are valid for arbitrary values of
$n_0$ and $a$. In the special case $a\gg n_0$ the results for $e_n(\tau)$,
$g_n(\tau)$ and $h_n(\tau)$ agree with those given in \cite{ON21}.

\section{Final remarks}
In conclusion, it has been demonstrated how the generating functions for density matrix
elements of the Jaynes-Cummings model with cavity damping may be written as sums over
eigenmodes with a fixed time dependence. The results (\ref{3.3}) and
(\ref{3.17})-(\ref{3.19}) contain coefficients (\ref{4.4}) and (\ref{4.21}) that are
adjusted to the initial conditions by means of the orthogonality relations (\ref{4.3}) and
(\ref{4.20}). The eigenmodes have been written in terms of generalised hypergeometric
functions.

The above analysis has been limited to a study of the generating functions for a suitable
subset of the density matrix elements, namely those with $m=n$, as discussed below
(\ref{2.2}). This is allowed as the complete collection of density matrix elements falls
apart in decoupled subsets, each with its own fixed value of $m-n$. For values $m-n\neq 0$
a similar analysis can be performed, although the details are somewhat more
complicated. In fact, one has to solve a set of four coupled equations instead of the
single equation (\ref{2.9}) and the three coupled ones given in (\ref{2.7}), (\ref{2.8})
and (\ref{2.10}).

\end{document}